\documentclass[aps,prl,twocolumn,superscriptaddress,amsfonts]{revtex4-1}

\usepackage{graphicx}
\usepackage{amsmath}
\usepackage{amssymb}
\usepackage{relsize}
\usepackage{xcolor}
\usepackage{filecontents}
\usepackage[exponent-product = \cdot]{siunitx}
\usepackage[colorlinks=true,citecolor=blue,linkcolor=magenta]{hyperref}
\hypersetup{
pdfauthor = {P. Maletinsky},
colorlinks = true, linkcolor = blue, urlcolor=blue, bookmarksnumbered =  true}

\newcommand{\bra}[1]{\langle #1|}
\newcommand{\ket}[1]{|#1\rangle}

\newcommand{\ketbra}[2]{\ket{#1}\!\bra{#2}}

\newcommand{\tr}{t_{\rm{r}}}
\newcommand{\del}{\partial}
\newcommand{\asin}{\sin^{-1}} %{\textrm{asin}}
\begin{document}

%============================
% TITLE OF PAPER
\title{Initialisation of single spin dressed states using shortcuts to adiabaticity}

\author{J.~K\"olbl}
\affiliation{Department of Physics, University of Basel, Basel 4056, Switzerland}
\author{A.~Barfuss}
\affiliation{Department of Physics, University of Basel, Basel 4056, Switzerland}
\author{M.~S.~Kasperczyk}
\affiliation{Department of Physics, University of Basel, Basel 4056, Switzerland}
\author{L.~Thiel}
\affiliation{Department of Physics, University of Basel, Basel 4056, Switzerland}
\author{A.~A.~Clerk}
\affiliation{Institute for Molecular Engineering, University of Chicago, Chicago, Illinois 60637, USA}
\author{H.~Ribeiro}
%\affiliation{Department of Physics, McGill University, Montr\'eal, Qu\'ebec H3A 2T8, Canada}
\affiliation{Max Planck Institute for the Science of Light, Erlangen 91058, Germany}
\author{P.~Maletinsky}
\email[]{patrick.maletinsky@unibas.ch}
\affiliation{Department of Physics, University of Basel, Basel 4056, Switzerland}

%\homepage[]{Your web page}
%\thanks{}

\date{\today}

%============================
% ABSTRACT
\begin{abstract}
We demonstrate the use of shortcuts to adiabaticity protocols for initialisation, readout, and coherent control of dressed states generated by closed-contour, coherent driving of a single spin.
Such dressed states have recently been shown to exhibit efficient coherence protection, beyond what their two-level counterparts can offer.
Our state transfer protocols yield a transfer fidelity of $\sim\!\SI{99.4(2)}{\%}$ while accelerating the transfer speed by a factor of \SI{2.6}{} compared to the adiabatic approach.
We show bi-directionality of the accelerated state transfer, which we employ for direct dressed state population readout after coherent manipulation in the dressed state manifold. 
Our results enable direct and efficient access to coherence-protected dressed states of individual spins and thereby offer attractive avenues for applications in quantum information processing or quantum sensing.
\end{abstract}

\maketitle

%============================
% INTRODUCTION
The pursuit of protocols for quantum sensing\,\cite{JonesEtAl2009, TaylorEtAl2008} and quantum information processing\,\cite{LossEtAl1998, Kane1998} builds on established techniques for initialising, coherently manipulating, and reading out quantum states, as extensively demonstrated in, e.g. trapped ions\,\cite{HomeEtAl2009, PoschingerEtAl2009}, solid state qubits\,\cite{ClarkeEtAl2008} and color centre spins\,\cite{HansonEtAl2008}.
Importantly, the involved quantum states need to be protected from decoherence\,\cite{NielsenEtAl2010}, which is primarily achieved by pulsed dynamical decoupling\,\cite{ViolaEtAl1999, Uhrig2007, DuEtAl2009}, a technique which suffers from drawbacks including experimental complexity and vulnerability to pulse errors. In contrast, `dressed states' generated by continuous driving of a quantum system yield efficient coherence protection\,\cite{WilsonEtAl2007, CaiEtAl2012a, GolterEtAl2014, TeissierEtAl2017}, even for comparatively weak driving fields \,\cite{BarfussEtAl2018}, in a robust, experimentally accessible way that is readily combined with quantum gates\,\cite{TimoneyEtAl2011, XuEtAl2012, LondonEtAl2013}.

A major bottleneck for further applications of such dressed states, however, is the difficulty in performing fast, high-fidelity initialisation into individual, well-defined dressed states. Up to now, such initialisation has focused on two-level systems and has mainly used adiabatic state transfer\,\cite{TimoneyEtAl2011, XuEtAl2012}, 
without detailed characterisation of the resulting fidelities. 
Adiabatic state transfer, however, suffers from a tradeoff between speed and fidelity: The initialisation must be slow to maintain fidelity, but fast enough to avoid decoherence during state transfer. For experimentally achievable driving field strengths, this tradeoff and the remaining sources of decoherence  form a key limitation to further advances in the use of dressed states in quantum information processing and sensing.   

\begin{figure}[tb]
	\centering
		\includegraphics{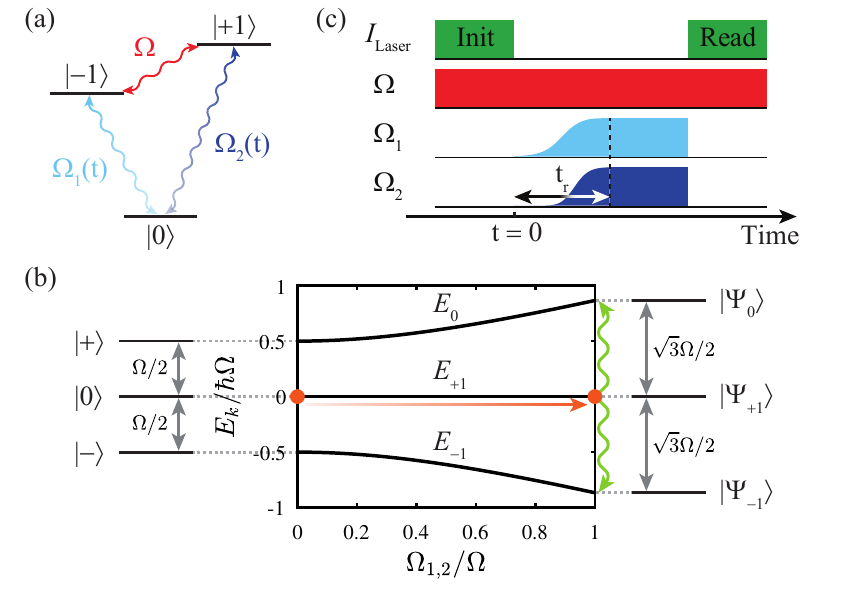}
	\caption{State transfer schematics. (a) Level scheme of the NV $S=1$ ground state spin with spin states $\ket{m_s}$ (magnetic quantum number $m_s=0,\pm 1$). All three spin transitions are individually and coherently addressable, either by MW magnetic fields ($\Omega_{1,2}(t)$, light and dark blue) or by a cantilever induced strain field ($\Omega$, red), enabling a CCD. 
	(b) Level schemes in the rotating frame for appropriate driving field phases [see text]. The initial system (left) comprises the states $\ket{0}$ and $\ket{\pm}$, with the $\ket{\pm}$ being equal admixtures of $\ket{\pm 1}$. Ramping the amplitudes of both MW fields causes a transfer to the final dressed states $\ket{\Psi_k}$ ($k=0,\pm 1$, right). We initialise the system in $\ket{0}$, which is for adiabatic ramping transferred to $\ket{\Psi_{+1}}$ (orange transition).
		(c) Pulse sequence employed for state transfer. Note that in general $\Omega_{1,2}(t)$ have different time dependencies.}
	\label{fig1}
\end{figure}

Here, we overcome these limitations with a twofold approach, where we employ recently developed protocols for `shortcuts to adiabaticity' (STA)\,\cite{DemirplakEtAl2003, Berry2009,CampoEtAl2010,ChenEtAl2010,ZhouEtAl2016} and apply them to the initialisation of three-level dressed states that exhibit efficient coherence protection, beyond what is offered by driven two-level systems\,\cite{BarfussEtAl2018}. 
Specifically, we focus on dressed states emerging from `closed-contour driving' (CCD) of a quantum three-level system\,\cite{BarfussEtAl2018} [Fig.\,\ref{fig1}(a)]. 
These dressed states stand out due to remarkable coherence properties and tunability through the phase of the involved driving fields\,\cite{BarfussEtAl2018}.
While dynamical decoupling by continuous driving has previously been demonstrated for electronic spins in diamond\,\cite{XuEtAl2012, CaiEtAl2012a,MacQuarrieEtAl2015,BarfussEtAl2015,BarfussEtAl2018}, STA have never been explored on such solid state spins in their ground state\,\cite{ZhouEtAl2016}, nor on the promising three-level dressed states we study here.
By combining STA and CCD, we establish an attractive, room-temperature platform for applications, e.g. in quantum sensing of high-frequency magnetic fields\,\cite{Joas2017,Stark2017a} on the nanoscale. 

We implement these concepts on individual Nitrogen-Vacancy (NV) electronic spins, which, due to their room-temperature operation and well-established methods for optical spin initialisation and readout\,\cite{GruberEtAl1997}, provide an attractive, solid-state platform for quantum technologies. 
The dressed states we study emerge from the $S = 1$ electronic spin ground state of the negatively charged NV centre, specifically from the eigenstates~$\ket{m_s}$ of the spin projection operator~$\hat{S}_z$ along the NV axis, with $m_s = 0,\pm 1$ being the corresponding spin quantum numbers [Fig.\,\ref{fig1}(a)]\,\cite{DohertyEtAl2013}.
To dress the NV spin states, we simultaneously and coherently drive all three available spin transitions, using microwave (MW) magnetic fields\,\cite{JelezkoEtAl2004} to drive the $\ket{0} \leftrightarrow \ket{\pm 1}$ transitions and time-varying strain fields\,\cite{BarfussEtAl2015,MacQuarrieEtAl2015} to drive the magnetic dipole-forbidden $\ket{-1} \leftrightarrow \ket{+1}$ transition 
[Fig.\,\ref{fig1}(a) and SOM].
The resulting CCD dressed states\,\cite{BarfussEtAl2018} offer superior coherence protection compared to alternative approaches, which rely on MW driving alone\,\cite{XuEtAl2012}. Specifically, CCD dressed states offer decoupling from magnetic field noise up to fourth order in field amplitude\,\cite{BarfussEtAl2018}, and for magnetometry, a more than twofold improvement in sensitivity and a 1000-fold increased sensing range [SOM] over previous work on dressed states\,\cite{XuEtAl2012}.

%============================
% SYSTEM
The CCD dressed states are best described in an appropriate rotating frame\,\cite{BuckleEtAl1986} where, under resonant driving of all three transitions, the system Hamiltonian reads
\begin{multline}
\hat{\mathcal{H}}_0(t) = \frac{\hbar}{2} \big( \Omega_1(t) \ketbra{-1}{0} + \Omega_2(t) \ketbra{+1}{0} + \\
+ \left. \Omega \, e^{i\Phi} \ketbra{-1}{+1} + \rm{H.\,c.} \right) \ ,
\label{eq.hamiltonian}
\end{multline}
with $\hbar$ being the reduced Planck constant.
The Hamiltonian $\hat{\mathcal{H}}_0(t)$ depends on the global driving phase $\Phi$ ($\Phi=\varphi+\varphi_1-\varphi_2$, where $\varphi,\varphi_{1}$, and $\varphi_{2}$ are the phases of the driving fields with Rabi frequecies $\Omega$, $\Omega_1$, and $\Omega_2$, respectively)\,\cite{BarfussEtAl2018}, which we tune to $\Phi = \pi/2$.
We choose this value of $\Phi$, as it allows for a straight-forward derivation of an analytical, purely real STA correction for our system. However, our method is applicable to arbitrary values of $\Phi$ using established numerical methods for determining the ensuing STA ramps [SOM].
For the case $\Phi = \pi/2$, the eigenstates of $\hat{\mathcal{H}}_0(t)$ prior to state transfer, i.e. for $\Omega_{1,2}(t) = 0$ and $\Omega \neq 0$, are $\ket{0}$ and $\ket{\pm} \equiv (\ket{-1} \mp i \ket{+1}) / \sqrt{2}$ [Fig.\,\ref{fig1}(b), left]. In contrast, the eigenstates of the final system, i.e. for $\Omega_{1,2}(\tr) = \Omega$, are given by\,\cite{BarfussEtAl2018}
\begin{align}
\ket{\Psi_k} = \frac{1}{\sqrt{3}} \left( e^{i\pi(1 - 4k)/6} \ket{-1} + \ket{0} + e^{-i\pi(1 - 4k)/6} \ket{+1} \right) \ ,
\label{eq.psi}
\end{align}
with $k=0,\pm 1$ [Fig.\,\ref{fig1}(b), right].
Thus our state transfer protocol consists of spin initialisation into $\ket{0}$ with $\Omega_{1,2}(t=0) = 0$, after which we apply suitable ramps $\Omega_{1,2}(t)$ to transfer into the dressed state basis with symmetric driving of all three transitions, i.e. $\Omega_{1,2}(t=\tr) = \Omega$ [Fig.\,\ref{fig1}(b)].

We study state transfer between the initial ($\ket{0}$, $\ket{\pm}$) and final states ($\ket{\Psi_k}$) under ambient conditions using an experimental setup described elsewhere\,\cite{BarfussEtAl2018} and by employing the pulse sequence shown in Fig.\,\ref{fig1}(c). A green laser pulse prepares the initial system in $\ket{\psi(t=0)} \equiv \ket{0}$. Then, we individually ramp the MW field amplitudes (with ramp time~$\tr$) to transfer $\ket{0}$ to the dressed state $\ket{\Psi_{+1}}$ [Fig.\,\ref{fig1}(b)]. After letting the system evolve in the presence of all three driving fields, we read out the population in $\ket{0}$ at time~$t$, $p_0(t) = \left|\left\langle 0 | \psi(t) \right\rangle \right|^2$,  using spin-dependent fluorescence. During the whole pulse sequence, the amplitude of the mechanical driving field is constant at $\Omega = 2\pi \cdot \SI{510}{kHz}$, while the mechanical oscillator is driven near resonance at \SI{5.868}{MHz} (implying $B_z= \SI{1.82}{G}$).

% FIGURE 1

\begin{figure}[tb]
	\centering
		\includegraphics{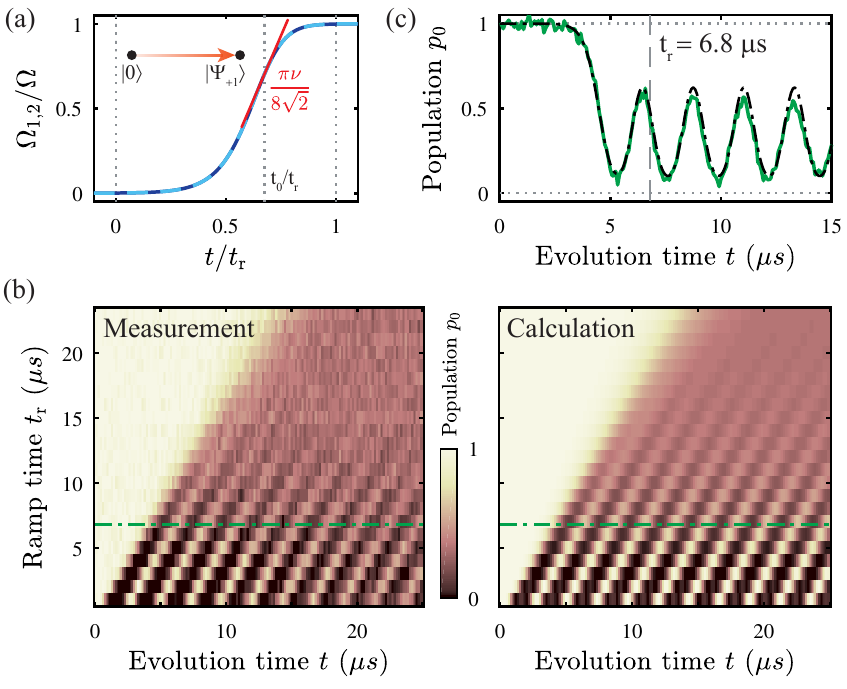}
	\caption{Adiabatic state transfer. (a) Experimentally employed ramps for the MW field amplitudes. Both MW fields are ramped simultaneously in an optimised STIRAP pulse shape [see text]. The slope parameter~$\nu$ is directly linked to the ramp time~$\tr$. (b) 2D~plots of $p_0 = \left| \left\langle 0 | \psi(t) \right\rangle \right|^2$ as a function of evolution time~$t$ and ramp time~$\tr$. The left plot shows experimental data, and the right plot shows theoretical calculations based on a fully coherent evolution. 
	For small $t_{\rm{r}}$ oscillations in the population indicate a non-adiabatic transfer, whereas for slower ramping of the MW fields the state transfer becomes adiabatic. 
	The green line indicates the location of the linecut presented in (c). (c) Linecut of measured population~$p_0$ (green) as function of evolution time~$t$ for $\nu = \Omega/2$ with corresponding calculation (black).}
	\label{fig2}
\end{figure}

%============================
% ADIABATIC STATE TRANSFER
To demonstrated state transfer into a dressed state, we first focus on an adiabatic protocol to benchmark our subsequent studies. 
Inspired by the `STIRAP' sequence developed for quantum-optical `$\Lambda$-systems'\,\cite{BergmannEtAl1998, VitanovEtAl2017}, we choose\,\cite{VasilevEtAl2009}
\begin{align}
\Omega_{1,2}(t) &= \Omega \cdot \sin\left( \theta(t) \right) \ ,
\label{eq.stirap1}
\end{align}
[Fig.\,\ref{fig2}(a)] with
\begin{align}
\theta(t) &= \frac{\pi}{2} \cdot \frac{1}{1+\exp(-\nu(t-t_0))}
\label{eq.stirap2}
\end{align}
being a Fermi function with time-shift $t_0 = \ln\left(\pi/(2 \,\asin(\varepsilon)) -1 \right) /\nu$ and free parameters $\varepsilon$ and $\nu$. Here, $\nu$ controls the slope of $\theta(t)$ at $t=t_0$ and is connected to the ramp time $\tr = t_0 - \ln\left( \pi/(2 \, \asin(1-\varepsilon)) - 1 \right) /\nu$, while $\varepsilon\ll1$ sets the amplitude of the ramp's unavoidable discontinuities at $t=0$ and $t=\tr$.
In all our experiments we use $\varepsilon = \SI{e-3}{}$, as this value is comparable to the estimated amplitude noise of our MW signals [SOM].

Figure\,\ref{fig2}(b) presents the time evolution of $p_0$ for several values of $\tr$. For fast ramping, i.e. small $\tr$, $p_0$ oscillates even for $t>\tr$; by increasing $\tr$, the amplitude of these oscillations reduces, until $p_0$ becomes time-independent with $p_0 \sim 1/3$. This marks a change from a non-adiabatic to an adiabatic transition with increasing $\tr$ [SOM]. Fast, non-adiabatic ramping results in a superposition of dressed states at the end of the ramp. During the subsequent time evolution each dressed state accumulates a dynamical phase, resulting in a beating (with frequency $\sqrt{3}\,\Omega/2$) in the measured population~$p_0$. Conversely, for larger $\tr$ we adiabatically prepare the single dressed state $\ket{\Psi_{+1}}$, where no such beating occurs and $p_0=\left| \left\langle 0 | \Psi_{+1} \right\rangle \right|^2 = 1/3$, as observed in the experiment.
We corroborate our experimental findings by calculating the time evolution $p_0(t)$ using Hamiltonian\,\eqref{eq.hamiltonian} and find excellent agreement with our data [Fig.\,\ref{fig2}(b)]. This agreement is additionally highlighted by the linecut in Fig.\,\ref{fig2}(c) taken in the non-adiabatic regime at $\tr = \SI{6.8}{\micro s}$ [green dashed line indicated in Fig.\,\ref{fig2}(b)]. 

% FIGURE 2

\begin{figure}[tb]
	\centering
		\includegraphics{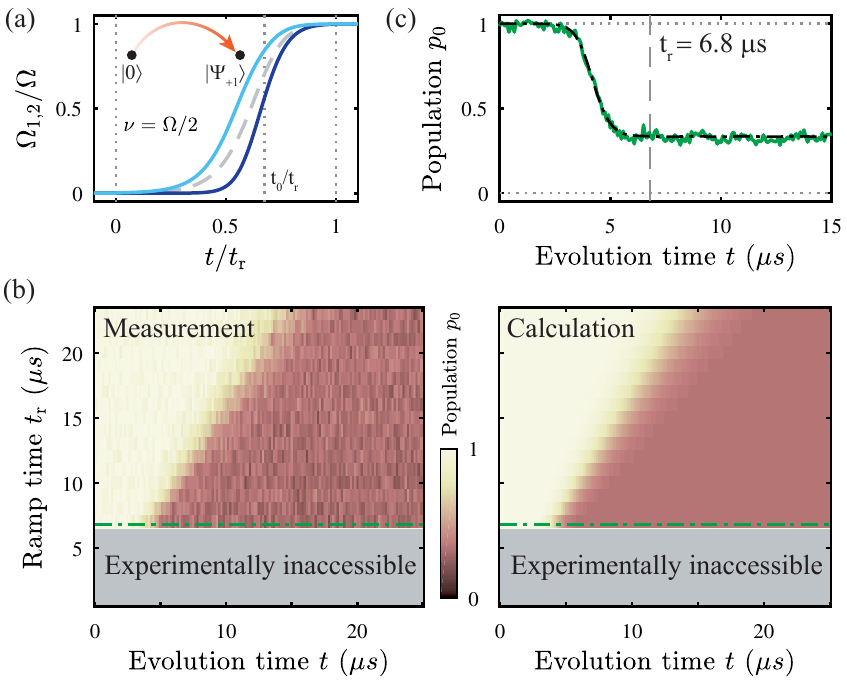}
	\caption{STA state transfer protocol. (a) Envelope of the optimised MW field amplitudes for $\nu=\Omega/2$ and $\Omega/2\pi = \SI{510}{kHz}$. Modifications of the adiabatic pulse shape (dashed) lead to altered and unequal ramps for the two MW fields ($\Omega_1$ light blue, $\Omega_2$ dark blue). (b) 2D~plots of $p_0$ as a function of evolution time~$t$ and ramp time~$t_r$. The left plot shows experimental data, and the right plot the theoretical calculations. We achieve state transfer with high fidelity independent of $t_{\rm{r}}$. The green line indicates the location of the linecut presented in (c) with ramp parameters at the experimental limit (border of grayed area). (c) Linecut of measured population~$p_0$ (green) as function of evolution time~$t$ for the same parameters as in Fig.\,\ref{fig2}(c) with corresponding calculation (black).}
	\label{fig3}
\end{figure}

%============================
% SUPERADIABATIC STATE TRANSFER
Having established adiabatic state transfer into the dressed state basis, we investigate STAs to speed up the initialisation procedure. Theoretical proposals provide various techniques for STA, including transitionless driving (TD)\,\citep{DemirplakEtAl2003, Berry2009} or the dressed state approach to STA\,\citep{BaksicEtAl2016}. All techniques harness non-adiabatic transitions by adding theoretically engineered corrections to the state transfer Hamiltonian. Adding a TD control results in the correction
\begin{align}
\hat{\mathcal{H}}_0(t) \rightarrow \hat{\mathcal{H}}_0(t) + i \hbar \left(\partial_t \,\hat{\mathcal{U}}^\dagger(t) \right) \hat{\mathcal{U}}(t) \ ,
\label{eq.td1}
\end{align}
with $\hat{\mathcal{U}}(t)$ being the transformation operator from $\{ \ket{m_s} \}$ into the adiabatic eigenstate basis\,\cite{DemirplakEtAl2003}.
We note that in our experiment, we can only implement the TD correction of Eq.\,(\ref{eq.td1}) for $\Phi=\pi/2$, where time reversal symmetry is maximally broken and the resulting TD correction is therefore purely real. An imaginary component would require control of the phase and amplitude of driving fields, which or mechanical-oscillator mediated strain drive cannot provide on the relevant timescales. For different values of $\Phi$, however, other STA ramps could be found using the dressed state formalism\,\cite{BaksicEtAl2016} [SOM]. Applying correction\,\eqref{eq.td1} to Hamiltonian\,\eqref{eq.hamiltonian} results in the modified MW pulse amplitudes
\begin{align}
\Omega_{1,2}(t) = \Omega \cdot \sin\left( \theta(t) \right) \pm 2 \cdot \frac{\cos(\theta(t)) \, \del_t \theta(t)}{2 - \cos(2 \theta(t))} \ ,
\label{eq.td2}
\end{align}
while keeping the phases of all fields constant.
Figure\,\ref{fig3}(a) shows the resulting MW pulse shapes for $\nu = \Omega/2$ and $\Omega = \SI{510}{kHz}$. Note that the TD approach provides different corrections for the two MW fields, such that both field amplitudes are ramped successively with different functional forms. 

Figure\,\ref{fig3}(b) depicts the experimental result of the state transfer when applying the TD corrected ramps. Independent of $\tr$, the time evolution of $p_0$ converges to $1/3$, which indicates perfect initialisation of a single dressed state, even for the fastest ramps, in striking agreement with the calculations. 
For TD driving, there exists a lower bound for $\tr$, below which the TD ramps lead to momentary driving field amplitudes either $<0$ or  $>\Omega$ [SOM]. For a fair comparison with adiabatic state transfer, we therefore exclude this parameter range from our study [grayed area in Fig.\,\ref{fig3}(b)].
The fastest possible state transfer corresponds to $\nu = \Omega/2$, resulting in $\tr = \SI{6.8}{\micro s}$ -- the value at which the data in Fig.\,\ref{fig3}(c) have been obtained. 

Figure\,\ref{fig2}(c) and Fig.\,\ref{fig3}(c) allow for a direct comparison of adiabatic and STA transfer protocols, since both measurements are recorded with the same set of experimental parameters. For the first approach, $p_0$ clearly indicates non-adiabatic errors in dressed state initialisation [Fig.\,\ref{fig2}(c)].
However, for the TD ramp, almost no oscillations in $p_0$ are visible, indicating excellent state transfer [Fig.\,\ref{fig3}(c)].
The remaining small oscillations are attributed to residual imperfections in the dressed state initialisation, which we discuss in the next paragraph. 
To achieve a transfer fidelity as determined by these residual oscillations, our calculations show that an adiabatic ramp of at least $\tr=\SI{17.6}{\micro s}$ would be required, which determines the speedup factor of $2.6$ we achieve for TD over adiabatic ramping for our given experimental parameters [SOM].

% FIGURE 3

\begin{figure}[tb]
	\centering
		\includegraphics{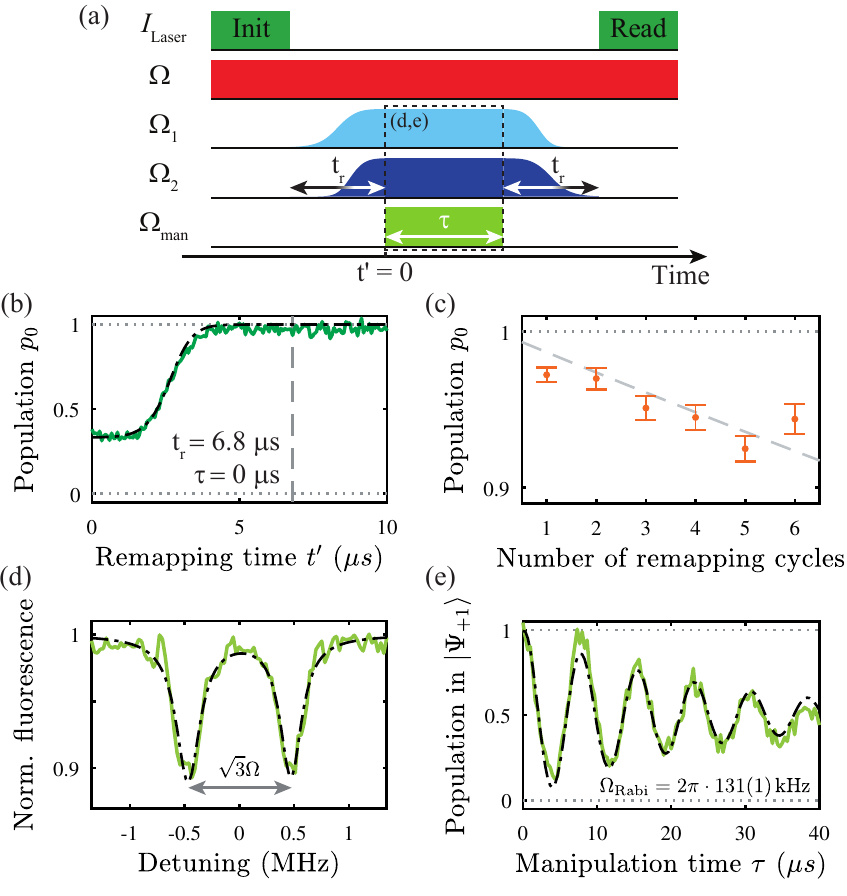}
		\caption{Transfer fidelity and dressed state characterisation. (a) Pulse sequence employed for bidirectional state transfer and manipulation of the dressed states. (b) Inverted state transfer to the initial system after the dressed state $\ket{\Psi_{+1}}$ was prepared. Measured population~$p_0$ (green) as function of the remapping time~$t^\prime$ for $\nu = \Omega/2$ with corresponding simulation (black). (c) Experimental transfer fidelity for various numbers of completed (re-)mapping cycles. The exponential fit to the data (dashed) yields a statistical transfer fidelity of \SI{99.4(2)}{\%}. (d) Transition frequencies of the dressed states after initialisation of $\ket{\Psi_{+1}}$ measured with an additional probe field~$\Omega_{\rm{man}}$, whose frequency is given relative to the $\ket{0}\leftrightarrow\ket{-1}$ transition of \SI{2.8672}{GHz}. (e) Rabi oscillations on a dressed state transition extracted from (d) with $\Omega_{\rm{Rabi}} = 2\pi \cdot \SI{131(1)}{kHz}$ and $T_{\rm{dec}} = \SI{24(3)}{\micro s}$.}
	\label{fig4}
\end{figure}

%============================
% REMAPPING AND STATE CHARACTERIZATION
To demonstrate reversibility and to verify that the TD protocol indeed results in initialisation of a single, pure dressed state, we reverse the state transfer and map from the dressed states back to the initial system. Specifically, we use the TD technique presented in Fig.\,\ref{fig3} to prepare the system in a single dressed state, and then use an inverted TD protocol (i.e. $t\rightarrow \tr -t$) to map back to the bare NV state $\ket{0}$ [Fig.\,\ref{fig4}(a)]. Figure\,\ref{fig4}(b) shows the time-evolution of $p_0$ as we apply the remapping protocol, where we set $\nu = \Omega/2$ (maximal ramping speed) for both directions. Clearly, almost all of the population in the dressed state returns to $\ket{0}$, thereby indicating coherent, reversible population transfer between undressed and dressed states. Additionally, such measurements allow us to quantify the efficiency of a single state transfer under the fair assumption that mapping in and mapping out yield the same transfer fidelity. We quantify the fidelity by repeatedly mapping in and out of the dressed state basis, with each set of one mapping in and one mapping out constituting a single `remapping cycle'. We vary the number of remapping cycles~$N$ and read out the population~$p_0$ at the end [Fig.\,\ref{fig4}(c)]. An exponential fit then yields the fidelity $F = \SI{99.4(2)}{\%}$ for a single transfer process. This transfer fidelity is experimentally limited by uncertainties in setting the global phase~$\Phi$, leakage of the MW signals, non-equal driving field amplitudes, and unwanted detunings of the driving fields. Although we calculate our ramps assuming equal driving amplitudes and zero detunings, violations of these assumptions are experimentally unavoidable, and the errors will generally fluctuate in time\,\cite{BarfussEtAl2018}. These factors are also responsible for the remaining, small oscillations in $p_0$ visible after state transfer in Fig.\,\ref{fig3}(c) [SOM].

Having shown efficient initialisation of a single, pure dressed state and subsequent dressed-state population readout, we next demonstrate coherent manipulation of dressed sates by performing electron spin resonance (ESR) and Rabi nutation measurements in the dressed state basis. For this, we apply an additional MW manipulation field of Rabi frequency $\Omega_{\rm{man}}$ in between the initialisation and remapping procedures [Fig.\,\ref{fig4}(a)]. By sweeping the frequency of this manipulation field (at a constant pulse duration $\tau=\SI{45}{\micro s}$) across the $\ket{0} \leftrightarrow \ket{-1}$ transition of the NV states, we observe two dips [Fig.\,\ref{fig4}(d)], corresponding to dressed state transitions at positive and negative frequencies in the rotating frame, i.e. at symmetric detunings around the bare $\ket{0} \leftrightarrow \ket{-1}$ transition frequency (note that the two possible transition from $\ket{\Psi_{+1}}$ to either $\ket{\Psi_{0}}$ or $\ket{\Psi_{-1}}$ [light green arrows in Fig.\,\ref{fig1}(b)] occur at the same frequencies and are therefore indistinguishable). Lastly, by resonant driving of the  dressed state transitions for varying durations $\tau$, we demonstrate coherent Rabi oscillations [Fig.\,\ref{fig4}(e)] and therefore coherent dressed state manipulation.

% FIGURE 4

%============================
% SUMMARY, CONCLUSION AND OUTLOOK
We have shown high-fidelity, reversible initialisation of individual dressed states in a CCD scheme using STA state transfer protocols, for which we demonstrated a more than twofold speedup over the adiabatic approach with state transfer fidelities $>99\%$. This performance is the direct result of our combination of STA (providing fast, high efficiency initialisation) and CCD dressed states (offering close to $100$-fold improvement in coherence times compared to the other continuous mechanical or MW driving schemes under similar conditions\,\cite{BarfussEtAl2018}). 
Our results provide a basis for future exploitation of dressed states, building on the coherent control of dressed states we have demonstrated. 
In particular, while the efficiency of coherence protection in CCD has been demonstrated recently\,\cite{BarfussEtAl2018}, details of additional dressed state dephasing mechanisms remain unknown and could be explored by employing noise spectroscopy\,\cite{Bar-Gill2012} and dynamical decoupling\,\cite{RyanEtAl2010, deLangeEtAl2010, NaydenovEtAl2011}, directly in the dressed state basis.
Owing to the prolonged dressed state coherence times over the bare spin states\,\cite{XuEtAl2012, LondonEtAl2013, MacQuarrieEtAl2015b, BarfussEtAl2018}, our technique could be used to efficiently store particular NV spin states by mapping to the dressed state basis\,\cite{SimonEtAl2010, SpechtEtAl2011} on timescales much longer than the coherence times of the bare NV states.
Lastly, owing to its versatility and stability, the experimental system we established here forms an attractive platform to implement and test novel state transfer protocols, which may emerge from future theoretical work.

%============================
% ACKNOWLEDGMENTS
\begin{acknowledgments}
We thank A. Stark for fruitful discussions and valuable input. We gratefully acknowledge financial support through the NCCR QSIT, a competence centre funded by the Swiss NSF, through the Swiss Nanoscience Institute, by the EU FP7 project ASTERIQS (grant \#$820394$) and through SNF Project Grant 169321.
\end{acknowledgments}

%============================
% BIBLIOGRAPHY
\bibliography{NV_CCI_STA}
\clearpage
\pagebreak
\newcommand{\beginsupplement}{%
	\setcounter{table}{0}
	\renewcommand{\thetable}{S\arabic{table}}%
	\setcounter{figure}{0}
	\renewcommand{\thefigure}{S\arabic{figure}}%
}

\beginsupplement
\onecolumngrid

%============================
% TITLE OF PAPER
%\appendix
\begin{center}
\large
\textbf{Supplementary Material for \\ ``Initialisation of single spin dressed states using shortcuts to adiabaticity''}
\end{center}

\normalsize

%============================
\section{Overview}
In the first part we describe our readout mechanism for the dressed state preparation measurements in detail and discuss the adiabatic criterion for the state transfer.
Next, we present a derivation of the transitionless driving (TD) correction. We then generalize our STA approach, where we calculate the required corrections for arbitrary phase values.
In the fourth part, we numerically calculate the Quantum Fisher information of our dressed states, from which we estimate their projection-noise-limited sensitivity.
In addition, we discuss the level structure of the NV and how we form a CCD scheme. 
Finally, we provide a detailed description of the MW field generation and control.

%============================ 
\section{Characterisation of state preparation process}
\label{sec:PreparationReadout}

Our experimental observations and theoretical calculations in Fig.\,2 and Fig.\,3 of the main text are based on the readout of the population
\begin{align}
p_0 = \left|\left\langle 0 | \psi(t) \right\rangle \right|^2 \ ,
\end{align}
where $\ket{\psi(t)}$ is the system's state after the time evolution $t$ [see Fig.\,1(c) of the main text].
The Hamiltonian determining the evolution of $\ket{\psi(t)}$ is $\mathcal{\hat{H}}_0(\tr)$ [see Eq.\,(1) of the main text], with $\Omega_{1,2}(\tr) = \Omega$, which leads to
\begin{align}
\ket{\psi(t)} = e^{-i(t-\tr) \mathcal{\hat{H}}_0(\tr)/\hbar} \ket{\psi(\tr)} \ .
\end{align}
If we prepare our system in a single dressed state, i.\,e. $\ket{\psi(\tr)} = \ket{\Psi_{+1}}$, the final state is
\begin{align}
\ket{\psi(t)} = e^{-i(t-\tr) \mathcal{\hat{H}}_0(\tr)/\hbar} \ket{\Psi_{+1}} = e^{-i(t-\tr) E_{+1} /\hbar} \ket{\Psi_{+1}} \ , 
\end{align}
with $E_{+1}$ being the eigenenergy corresponding to $\ket{\Psi_{+1}}$, which characterises the dynamical phase the eigenstate accumulates during its evolution.
As our readout state is $\ket{0} = (\ket{\Psi_{-1}} + \ket{\Psi_0} + \ket{\Psi_{+1}})/\sqrt{3}$, the measured population $p_0$ can be expressed as
\begin{align}
p_0 = \frac{1}{3} \left|\sum_{k=0,\pm 1} \left\langle \Psi_k \left| e^{-i(t-\tr) E_{+1} /\hbar} \right| \Psi_{+1} \right\rangle \right|^2 = \frac{1}{3} \ .
\end{align}
Thus, for a perfect state transfer into the state $\ket{\Psi_{+1}}$, the measured population is time independent with a value of 1/3.
The same holds for the other two dressed states, $\ket{\Psi_{0}}$ and $\ket{\Psi_{-1}}$. \\

If, however, we do not prepare a single dressed state, but rather a mixture of dressed states, we start the time evolution in $\ket{\psi(\tr)} = \sum_{j=0,\pm 1} c_j \ket{\Psi_{j}}$, with $c_j \in \mathbb{C}$ and $\sum_j |c_j|^2 = 1$. 
In this case, the final state of the evolution is given by
\begin{align}
\ket{\psi(t)} = e^{-i(t-\tr) \mathcal{\hat{H}}_0(\tr)/\hbar} \sum_{j=0,\pm 1} c_j \ket{\Psi_{j}} = \sum_{j=0,\pm 1} c_j \, e^{-i(t-\tr) E_j /\hbar} \ket{\Psi_{j}} \ ,
\end{align}
as each dressed state $\ket{\Psi_j}$ accumulates a dynamical phase corresponding to its eigenenergy $E_j$ ($j=0,\pm 1$). During readout we then measure
\begin{align}
p_0 = \frac{1}{3} \left|\sum_{j,k=0,\pm 1} \left\langle \Psi_k \left| c_j \, e^{-i(t-\tr) E_{j} /\hbar} \right| \Psi_{j} \right\rangle \right|^2 = \frac{1}{3} \left|\sum_{j=0,\pm 1} c_j\, e^{-i(t-\tr) E_j /\hbar} \right|^2  \ ,
\end{align}
resulting in a time dependence of $p_0$ characterized by a beating of the transition frequencies of the dressed states with amplitudes depending on the weighting factors $c_j$. \\

With that, we can explain the observed transition from an oscillatory to a constant time evolution with increasing ramp time $\tr$ in Fig.\,2(b) of the main text. For fast ramping, i.\,e. small $\tr$, the preparation is non-adiabatic, and we therefore prepare a mixture of dressed states, resulting in an oscillatory $p_0$, which we observe experimentally. Increasing $\tr$ decreases the amplitude of these oscillations, indicating that one weighting factor becomes dominant, so that the dressed state mixing is reduced. If we finally reach the adiabatic regime, only a single dressed state is prepared resulting in a time independent evolution of the measured population. \\

To quantify the transition from the non-adiabatic to the adiabatic regime we compare the energy separation of the instantaneous adiabatic eigenstates with their mutual coupling. Therefore, we calculate Hamiltonian~$\mathcal{\hat{H}}_0(t)$ [see Eq.\,(1) of the main text] in the adiabatic basis\,\cite{DemirplakEtAl2003}:
\begin{align}
\mathcal{\hat{H}}_{0,\rm{ad}}(t) = \mathcal{\hat{U}}(t) \, \mathcal{\hat{H}}_0(t) \, \mathcal{\hat{U}}^\dagger(t) - i \hbar \; \mathcal{\hat{U}}(t) \, \partial_t \mathcal{\hat{U}}^\dagger(t) \ .
\label{fml.AdiabaticHamiltonian}
\end{align}
Here, the time dependent unitary transformation $\mathcal{\hat{U}}(t)$ is the transformation operator from the $\{ \ket{m_s} \}$ basis to the adiabatic basis states (i.\,e. the instantaneous eigenvectors of $\mathcal{\hat{H}}_0(t)$).
We obtain $\mathcal{\hat{U}}(t)$ from its inverse $\mathcal{\hat{U}}^\dagger(t)$, whose columns contain the time dependent, normalised adiabatic eigenvectors. \\

Considering the $\{ \ket{-1},\ket{0},\ket{+1} \}$ basis we find
\begin{align}
\hat{\mathcal{H}}_0(t) = \frac{\hbar \, \Omega}{2} \begin{pmatrix} 0 & \sin(\theta(t)) & i \\ \sin(\theta(t)) & 0 & \sin(\theta(t)) \\ -i & \sin(\theta(t)) & 0 \end{pmatrix} \ ,
\label{fml.hamiltonian}
\end{align}
and 
\begin{align}
\mathcal{\hat{U}}^\dagger(t) =
\begin{pmatrix}
\frac{1}{2} + \frac{i}{2 \sqrt{2-\cos(2\theta(t))}} & -\frac{\sin(\theta(t))}{\sqrt{2-\cos(2\theta(t))}} & \frac{1}{2} - \frac{i}{2 \sqrt{2-\cos(2\theta(t))}} \\
\frac{\sin(\theta(t))}{\sqrt{2-\cos(2\theta(t))}} & -\frac{i}{\sqrt{2-\cos(2\theta(t))}} & -\frac{\sin(\theta(t))}{\sqrt{2-\cos(2\theta(t))}} \\
\frac{1}{2} - \frac{i}{2 \sqrt{2-\cos(2\theta(t))}} & \frac{\sin(\theta(t))}{\sqrt{2-\cos(2\theta(t))}} & \frac{1}{2} + \frac{i}{2 \sqrt{2-\cos(2\theta(t))}}
\end{pmatrix} \ .
\label{fml.Udagger}
\end{align}
Taking the Hermitian conjugate of Eq.\,\eqref{fml.Udagger} and inserting into Eq.\,\eqref{fml.AdiabaticHamiltonian} yields
\begin{align}
\mathcal{\hat{H}}_{0,\rm{ad}}(t) = \hbar
\begin{pmatrix}
\frac{\Omega}{2} \sqrt{2-\cos(2\theta(t))} & \frac{\cos(\theta(t)) \, \partial_t \theta(t)}{2- \cos(2\theta(t))} & 0 \\
\frac{\cos(\theta(t)) \, \partial_t \theta(t)}{2- \cos(2\theta(t))} & 0 & - \frac{\cos(\theta(t)) \, \partial_t \theta(t)}{2- \cos(2\theta(t))} \\
0 & - \frac{\cos(\theta(t)) \, \partial_t \theta(t)}{2-\cos(2\theta(t))} & -\frac{\Omega}{2} \sqrt{2-\cos(2\theta(t))}
\end{pmatrix} \ .
\label{fml.Hadiab}
\end{align}
In order to realise an adiabatic state transfer, the coupling between two instantaneous adiabatic eigenstates has to be much smaller than their energy separation, i.\,e. we have to compare the skew diagonal elements of Eq.\,\eqref{fml.Hadiab} with the separation of the diagonal elements. Thus, in our case the adiabatic criterion reads
\begin{align}
\frac{2 \cos(\theta(t)) \, \partial_t \theta(t)}{\Omega \big( 2- \cos(2\theta(t)) \big)^{3/2}} \ll 1 \ .
\end{align}
To estimate the transition from satisfying to violating the adiabatic criterion, we choose an upper limit for the right-hand side of $1/10$ for all times~$t$, i.\,e.
\begin{align}
\max \left\{ \frac{2 \cos(\theta(t)) \, \partial_t \theta(t)}{\Omega \big( 2- \cos(2\theta(t)) \big)^{3/2}} \right\} \leq \frac{1}{10} \ .
\end{align}
Thus in this paper we call a ramp time $\tr$ which satisfies this inequality adiabatic, and we call a ramp time that violates this inequality non-adiabatic. With this definition, we find that $\tr = \SI{12.9}{\micro s}$ is the critical ramp time for an adiabatic transition. This corresponds to a theoretical fidelity $\mathcal{F} = |\langle \Psi_+ | \psi(\tr) \rangle|^2$ of $(1-\SI{3.2e-3}{})$, which is lower than the asymptotic fidelity of $(1-\SI{2e-6}{})$ by \SI{0.35}{\%} percent.

%============================ 
\section{Derivation of the TD correction}
\label{sec:TDcorrection}

\begin{figure}[tb]
	\centering
		\includegraphics{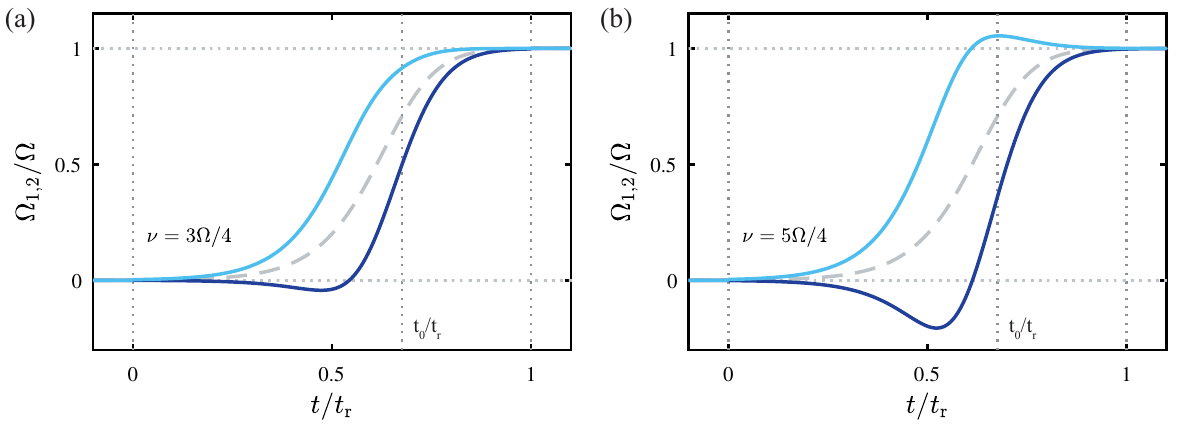}
		\caption{Theoretically calculated TD corrected MW ramp shapes for $\nu > \Omega/2$. (a) For $\nu = 3\Omega/4$ the envelope of the optimised ramp has negative amplitude values, which cannot be implemented experimentally. (b) Increasing $\nu$ further additionally leads to overshoots, e.\,g. here displayed for $\nu = 5\Omega/4$.}
	\label{fig2som}
\end{figure}

To calculate the TD correction of our microwave (MW) pulses stated in Eq.\,(6) of the main text, we consider the Hamiltonian $\mathcal{\hat{H}}_{0,\rm{ad}}(t)$ in the adiabatic basis [see Eq.\,\eqref{fml.Hadiab}]. In order to get rid of the off-diagonal elements, i.\,e. to ensure a transfer on a single adiabatic eigenstate, additional control fields expressed by a control Hamiltonian $\mathcal{\hat{H}}_{\rm{TD}}(t)$ have to be added to the system. As could be anticipated by inspecting Eq.\,\eqref{fml.AdiabaticHamiltonian}, a general expression for the TD control Hamiltonian in the $\{ \ket{m_s} \}$ basis reads\,\cite{DemirplakEtAl2003}:
\begin{align}
\mathcal{\hat{H}}_{\rm{TD}}(t) = i \hbar \left( \partial_t \mathcal{\hat{U}}^\dagger(t) \right) \mathcal{\hat{U}}(t) \ .
\label{fml.TDHamiltonian}
\end{align}
Here, $\mathcal{\hat{U}}^\dagger(t)$ is the unitary operator given in Eq.\,\eqref{fml.Udagger}. Inserting the corresponding expressions into Eq.\,\eqref{fml.TDHamiltonian} we yield in the $\{ \ket{-1}, \ket{0}, \ket{+1} \}$ basis:
\begin{align}
\mathcal{\hat{H}}_{\rm{TD}}(t) = \hbar 
\begin{pmatrix}
0 & \frac{\cos(\theta(t)) \, \partial_t \theta(t)}{2- \cos(2\theta(t))} & 0 \\
\frac{\cos(\theta(t)) \, \partial_t \theta(t)}{2- \cos(2\theta(t))} & 0 & -\frac{\cos(\theta(t)) \, \partial_t \theta(t)}{2-\cos(2\theta(t))} \\
0 & - \frac{\cos(\theta(t)) \, \partial_t \theta(t)}{2- \cos(2\theta(t))} & 0
\end{pmatrix} \ .
\end{align}
After adding this Hamiltonian to the initial Hamiltonian $\mathcal{\hat{H}}_0(t)$ from Eq.\,\eqref{fml.hamiltonian}, we find that the TD corrected MW pulses are
\begin{align}
\Omega_{1,(2)}(t) = \Omega \cdot \sin(\theta(t)) \, \substack{+\\[-0.25em]\mathsmaller{(-)}}\; 2 \cdot \frac{\cos(\theta(t)) \, \partial_t \theta(t)}{2- \cos(2\theta(t))} \ .
\end{align}
In the experiment we can easily realize TD corrected MW pulses with $0 \leq \Omega_{1,(2)}(t) \leq \Omega$, which require the parameter~$\nu$ to be $\nu \leq \Omega/2$. If we choose $\nu > \Omega/2$, however, the TD corrected ramps would lead to negative amplitude values, which we cannot implement experimentally, since flipping the phase of the signal generation fields does not influence the final driving field phases (see later section for details on the MW pulse generation). Figure\,\ref{fig2som}(a) displays an example for such a ramp for $\nu = 3\Omega/4$. For even higher values of $\nu$ some amplitudes additionally overshoot the steady state driving amplitude (see Fig.\,\ref{fig2som}(b) for $\nu = 5\Omega/4$). \\

% FIGURE SOM

\begin{figure}[h]
	\centering
		\includegraphics{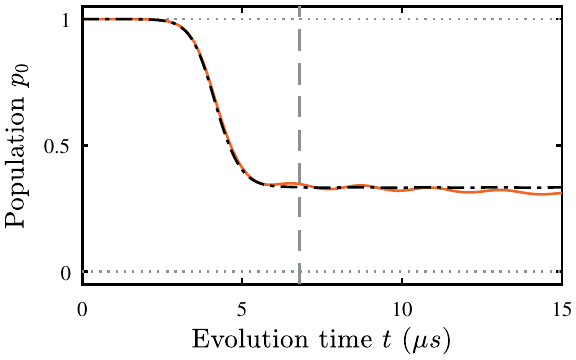}
		\caption{Simulated influence of experimental uncertainties on the state preparation process. While the unperturbed system shows no oscillations after the state transfer (dash-dotted line in black), including experimental uncertainties reveals the experimentally observed oscillations (solid line in orange). Parameters: $\nu = \Omega/2$, $\Omega/2\pi = \SI{510}{kHz}$, $\Omega_1^{\rm{max}}/2\pi = \SI{520}{kHz}$, $\Omega_2^{\rm{max}}/2\pi = \SI{500}{kHz}$ (i.\,e. \SI{2}{\%} opposing deviation for both MW fields), $\Phi = 0.495 \cdot \pi/2$. The perturbed simulation is averaged over $N_{\rm{avg}} = 100$ times, with each iteration using a different MW detuning drawn from a zero-mean Gaussian distribution characterized by $\sigma_{T_2^*}/2\pi = \SI{107}{kHz}$.}
	\label{figstatprepsimulation}
\end{figure}

As mentioned in the main text, the theoretically derived TD corrections do not account for the unavoidable experimental uncertainties that cause the residual small oscillations in $p_0$ after the state transfer in Figure\,3(c). Most importantly, our measurements are affected by slow fluctuations in the magnetic environment of the NV caused by nearby nuclear spins ($^{14}$N or $^{13}$C) and by uncertainties in setting the driving field parameters. The magnetic fluctuations are characterized by a zero-mean Gaussian distribution of the MW field detunings with a width of $\sigma_{T_2^*}/2\pi = 1/(\sqrt{2}\pi T_2^*) = \SI{107}{kHz}$\,\cite{JamonneauEtAl2016}, where $T_2^* = \SI{2.1(1)}{\micro s}$ is the coherence time determined through a Ramsey experiment.
The detuning is typically constant during a single measurement,but changes between measurements, i.e. the timescale of the fluctuations is long compared to a single measurement, but far shorter than the total measurement time.
The uncertainties in setting the driving field parameters mostly affect the MW driving strengths.
We measure the driving strengths by driving Rabi oscillations on each of the MW transitions of the bare NV and extracting the Rabi frequency by fitting with an exponentially decaying single sinusoid. But due to fluctuations in, for example, the sample-antenna separation, as well as the microwave amplifier, the measured Rabi frequencies show relative deviations of up to \SI{2}{\%} for the same applied microwave power ($\sim \SI{4}{dBm}$ from the SRS microwave generator in our case). Moreover, setting the global phase value exactly to $\Phi= \pi / 2$ has experimental limitations. We determine the corresponding phase value by sweeping the mechanical phase with finite sampling rate while maintaining the MW fields constant [see later] and fitting the averaged linecuts of the resulting interference pattern [compare to\,\cite{BarfussEtAl2018}] for evolution times between \SI{1}{\micro s} and \SI{1.7}{\micro s}. This allows us to determine the value of the global phase with a $2\sigma$ uncertainty of \ang{0.9}. Figure\,\ref{figstatprepsimulation} shows the simulated population $p_0$ as a function of evolution time $t$ averaged over $N_{\rm{avg}} = 100$ normally distributed MW detunings, while also including other experimental uncertainties in a worst-case scenario (i.e. maximum global phase error, and maximum deviation in drive strengths).
The time evolution clearly shows oscillations after the state transfer, similar to those observed in the experiment.

We note that there are additional sources for experimental uncertainties that we neglect in our simulation, e.\,g. the feedthrough of the MW signals (which leads to non-vanishing MW amplitudes at the beginning of the state transfer), amplitude and frequency noise of the driving fields, as well as fluctuations in the zero-field splitting of the NV induced by variations in temperature or environmental strain or electric fields.

%============================ 
\section{Finding STAs for arbitrary phase values}
\label{sec:dressedSTA}

STAs for arbitrary phase values $\Phi$ can be found for our current setup (no time-dependent phase control on the MW fields and on the amplitude of the mechanical drive) using the dressed state approach introduced in Ref.\,\cite{BaksicEtAl2016}.
The basic idea is to modify the control fields of the Hamiltonian $\hat{\mathcal{H}}(t)$ [see Eq.\,(1) of the main text], so that the state transfer is realized without errors even when the protocol time is not long compared to the instantaneous adiabatic gap. The modification of the control fields can be described as a modification of the initial Hamiltonian by adding an extra control term, i.\,e. $\hat{\mathcal{H}}(t) \to \hat{\mathcal{H}}_\mathrm{mod}(t) = \hat{\mathcal{H}}(t) + \hat{\mathcal{W}}(t)$. We can parametrize $\hat{\mathcal{W}}(t)$ as 
\begin{align} 
	\hat{\mathcal{W}}(t) = W_1(t) \ketbra{-1}{0} + W_2(t) \ketbra{+1}{0} + \mathrm{H.c.} \ , 
	\label{eq:CtrlFields}
\end{align}
where for our experimental setup $W_1(t)$ and $W_2(t)$ are real valued functions. We, however, note that if time-dependent phase control is available, then $W_1(t)$ and $W_2(t)$ can be complex valued. We also note that Eq.\,\eqref{eq:CtrlFields} can be further generalized to take into account possible time-dependent control of the mechanical drive. 

There are an infinity of possible $\hat{\mathcal{W}}(t)$, each of them being associated to a specific dressing of the instantaneous eigenstate used to realized the adiabatic state preparation. We choose $\hat{\mathcal{W}}(t)$ such that in the dressed adiabatic frame the evolution of the dressed state used for state preparation is trivial, i.\,e. one remains in this particular dressed state for the whole duration of the protocol. In our case the dressing transformation can be generated by the unitary operator
\begin{align}
	\hat{\mathcal{U}}_\mathrm{dress}(t) = \prod_j e^{i \varphi_j(t) \hat{\lambda}_j} \ ,
	\label{eq:UDress}
\end{align}
where $\hat{\lambda}_j$ are the generators of $\mathrm{SU}(3)$ and $\varphi_j(t)$ must obey the  conditions $\varphi_j(0) = \varphi_j(t_\mathrm{r}) = 0$ to ensure that the dressed states correspond to the adiabatic states of Eq.\,(1) of the main text at the beginning and end of the protocol. This ensures that one starts and ends in the desired state.

To make things concrete let us explicitly write the equations that help us determine $\hat{\mathcal{W}}(t)$. We start by finding the adiabatic states (instantaneous eigenstates) of Eq.\,(1) of the main text. We need to solve the eigenvalue problem 
\begin{align}
	\hat{\mathcal{H}}(t) \ket{\psi_k(t)} = E_k(t) \ket{\psi_k(t)} \ .
	\label{eq:InstEigEAndEigVec}
\end{align}
We find the instantaneous eigenenergies 
\begin{align}
	E_k(t) = \frac{\Omega}{\sqrt{3}}\sqrt{1 + 2 \Omega(t)^2} \, \cos \left[ \frac{1}{3} \arccos \left(3 \sqrt{3} \frac{\cos(\Phi) \, \Omega(t)^2}{(1 + 2 \Omega(t)^2)^{\frac{3}{2}}} \right) - \frac{2 \pi}{3} k\right] \ ,
	\label{eq:InstEigE}
\end{align}
where $\Omega(t) = \Omega_1(t) = \Omega_2(t) = \Omega \sin(\theta(t))$. The associated eigenvectors are given by 
\begin{align}
		\ket{\psi_k(t)} &= \frac{1}{N_k(t)} \left[ \Omega(t) \left( \Omega e^{-i \Phi} + 2 E_k(t) \right) \ket{+ 1} - \left( \Omega^2 - 4 E_k(t) \right) \ket{0} \right. \notag \\
		&\phantom{={}} \left. + \Omega(t) \left( \Omega e^{i \Phi} + 2 E_k(t) \right) \ket{- 1} \right] \ ,
	\label{eq:InstEigVec}
\end{align}
with $N_k(t) = \sqrt{ (\Omega^2 - 4 E_k^2(t))^2 + 2 \Omega^2(t) \, [\Omega^2 + 4 E_k(t) \, (\Omega \cos(\Phi) + E_k(t))]}$ the normalization factor. To transform Eq.\,(1) of the main text to the adiabatic frame (the instantaneous eigenstate basis), we define the frame-change operator given by the time-dependent unitary 
\begin{align}
	\hat{\mathcal{U}}(t) = \sum_{k=\pm 1, 0} \ketbra{\psi_k(t)}{\psi_k} \ ,
	\label{eq:SAd}
\end{align}
which diagonalizes $\hat{\mathcal{H}} (t)$ at each instant in time. \\

We can now express the dressed adiabatic Hamiltonian as 
\begin{align}
		\hat{\mathcal{H}}_\mathrm{dress}(t) &= \hat{\mathcal{U}}_\mathrm{dress}^\dag (t) \left[\hat{\mathcal{U}}^\dag(t) \hat{\mathcal{H}}(t) \hat{\mathcal{U}}(t) + i \left( \partial_t  \hat{\mathcal{U}}^\dag(t) \right) \hat{\mathcal{U}}(t) + \hat{\mathcal{U}}^\dag(t)  \hat{\mathcal{W}}(t) \hat{\mathcal{U}}\right] \hat{\mathcal{U}}_\mathrm{dress}(t) \notag \\
	&\phantom{={}} + i \left( \partial_t  \hat{\mathcal{U}}_\mathrm{dress}^\dag(t) \right) \hat{\mathcal{U}}_\mathrm{dress}(t)
	\label{eq:HModAd}
\end{align}
and the dressed adiabatic states as 
\begin{align}
	\ket{\tilde{\psi}_k(t)} =  \hat{\mathcal{U}}_\mathrm{dress}^\dag(t)  \ket{\psi_k} \ .
	\label{eq:DressState}
\end{align}
Given that the adiabatic state transfer is realized through $\ket{\psi_{+1}(t)}$, we ask that $\ket{\tilde{\psi}_{+1}(t)}$ is decoupled from the other dressed states, which ensures the desired trivial dynamics. This condition can be written as 
\begin{align}
	\bra{\tilde{\psi}_{+1}} \hat{\mathcal{H}}_\mathrm{dress}(t) \ket{\tilde{\psi}_{j}} = 0 \qquad j=0,-1 \ ,
	\label{eq:CondW}
\end{align}
where the states $\ket{\tilde{\psi}_k}$ are time-independent since they are expressed in the frame defined by $\hat{\mathcal{U}}_\mathrm{dress}^\dag(t)$. Solving this system of equation for a chosen dressing gives $\hat{\mathcal{W}}(t)$. \\

For instance, for $\Phi=0$ the TD correction only has purely imaginary matrix elements and the SATD correction (see Ref.\,\cite{BaksicEtAl2016}) requires controlling the detunings of all states and the amplitude of the mechanical drive. None of these corrections can be implemented with our current setup. However, using the dressed state method one can find a STA that respects the constraints of our system. We choose a dressing of the form
\begin{align}
	\hat{\mathcal{U}}_{\mathrm{dress},\Phi=0}(t) = e^{-i \frac{\beta(t)}{2} (\ketbra{\psi_0}{\psi_{+1}} + \mathrm{H.c.})} \ .
	\label{eq:DressPhi0}
\end{align}
Using Eq.\,\eqref{eq:CondW}, we find that the equation $\bra{\tilde{\psi}_{+1}} \hat{\mathcal{H}}_\mathrm{dress}(t) \ket{\tilde{\psi}_{-1}} = 0$ can be fulfilled by choosing $W_1 (t) = W_2 (t)$ while $\bra{\tilde{\psi}_{+1}} \hat{\mathcal{H}}_\mathrm{dress}(t) \ket{\tilde{\psi}_0} = 0$ can be fulfilled by solving 
\begin{align}
	\beta (t) &= \arctan\left[\frac{4 \sqrt{2} \, \cos(\theta (t)) \,  \dot{\theta} (t)}{\sqrt{1+8 \sin^2(\theta (t))} \, \left[\Omega \left(1+8 \sin^2(\theta (t)) + 16 W_1(t) \, \sin(\theta(t))\right)\right]}\right] 
	\label{eq:solDressBeta} \\
	0 &= \frac{\sqrt{2} \, W_1(t)}{\sqrt{1+8 \sin^2(\theta (t))}} - \frac{\dot{\beta}(t)}{2} \ .
	\label{eq:SolDressW}
\end{align}
Inserting Eq.~\eqref{eq:solDressBeta} into Eq.~\eqref{eq:SolDressW} one gets a differential equation for $W_1 (t)$. 

%============================ 
\section{Quantum Fisher Information of Dressed States}
\label{sec:QFI}

To estimate how suitable our CCI three-level dressed states are for sensing, we calculate their quantum Fisher information (QFI) and compare it to the QFI of three-level dressed states formed with two microwave driving fields\,\cite{XuEtAl2012}.
Using the QFI we estimate the sensitivity of these states and show that our CCI dressed states can achieve projection-noise-limited sensitivities down to $\approx \SI{2.75}{nT/\sqrt{Hz}}$, compared to $\approx \SI{6.43}{nT/\sqrt{Hz}}$ for the two microwave (2MW) dressed states (see below for formulae giving the driving Hamiltonian and resulting 2MW dressed states).
Additionally, we show that our CCI dressed states can sense a $\sim\!1000$ times broader range of magnetic field strengths.

We consider the case in which we use our states to sense a static magnetic field $B_z$. We are therefore interested in the QFI $J(B_z)$ of our states as a function of $B_z$. The QFI is defined as
\begin{align}
J(B_z) = \mathrm{Tr} \lbrace \left( \partial_{B_z} \hat{\rho} \right) \hat{L}_{B_z} \rbrace \ ,
\end{align}
\noindent where $\partial_{B_z} \hat{\rho}$ indicates the derivative of the density matrix $\hat{\rho}$ with respect to $B_z$\,\cite{Chaudhry2014}.
The operator $\hat{L}_{B_z}$ is given by
\begin{align}
\hat{L}_{B_z} = 2 \sum_{k,\ell} \frac{\bra{ \phi_k } \partial_{B_z} \hat{\rho} \ket{ \phi_\ell }}{\rho_k + \rho_\ell} \ket{ \phi_k } \bra{\phi_\ell } \, ,
\end{align}
\noindent where the sum is over all eigenstates $\lvert \phi_k \rangle$ (with eigenvalues $\rho_k$) of $\hat{\rho}$, such that $\rho_k + \rho_\ell \neq 0$\,\cite{Chaudhry2014}. The QFI sets a lower bound on the variance $\delta B_z ^2$ of our estimate of $B_z$ through the quantum Cram\'{e}r-Rao bound
\begin{align}
\delta B_z ^2 \geq \frac{1}{J(B_z)}
\end{align}
\noindent for an unbiased estimator of a single measurement of $B_z$\,\cite{Chaudhry2014}.
We assume that we can find the optimal estimator such that the equality obtains.
The sensitivity is broadly defined as the weakest magnetic field that can be sensed while still achieving an signal-to-noise ratio (SNR) of 1 after $N$ measurements\,\cite{DegenEtAl2017}.
We define the SNR as 
\begin{align}
\mathrm{SNR} = \frac{\sqrt{N} B_z}{\delta B_z} \ ,
\end{align}
\noindent so that an SNR of 1 corresponds to $B_\mathrm{min} = \delta B_z / \sqrt{N}$.
The optimal evolution time for a single measurement of $B_z$ is approximately the coherence time of the sensing state, such that for a total measurement time $T$ we can make $N = T / T_{\mathrm{coh}}$ measurements (neglecting state preparation and measurement overhead times) \cite{DegenEtAl2017}.
In terms of $T_{\mathrm{coh}}$ and $J(B_z)$, we therefore have
\begin{align}
B_{\mathrm{min}} = \frac{\sqrt{T_\mathrm{coh}}}{\sqrt{T} \sqrt{J(B_\mathrm{min}; T_\mathrm{coh})}} \ ,
\label{fml:Bmin}
\end{align}
\noindent where we have indicated that we evaluate $J(B_z)$ at the coherence time $T_\mathrm{coh}$ and the minimum field $B_\mathrm{min}$ that can be sensed with an SNR of 1.
The sensitivity $S$ (which characterizes the weakest measurable magnetic field, in units of $\mathrm{T / \sqrt{Hz}}$) is thus given by
\begin{align}
S = \frac{\sqrt{T_\mathrm{coh}}}{\sqrt{J(B_\mathrm{min}; T_\mathrm{coh})}} \ .
\label{fml:Sensitivity}
\end{align}
\noindent To determine the sensitivity of our states, we therefore need to know their coherence times and their QFI. Since the analytic expression for $J(B_z; T_\mathrm{coh})$ is unknown, in practice we numerically evaluate Eq.\,\eqref{fml:Bmin} to approximate $B_\mathrm{min}$, i.\,e. we take $B_\mathrm{min}$ to be the value of $B_z$ that gives $\mathrm{min} \lbrace \lvert 1 - B_z \sqrt{J(B_z; T_\mathrm{coh})} / \sqrt{T_\mathrm{coh}} \rvert \rbrace$. Once we find $B_\mathrm{min}$, we can evaluate Eq.~\,\eqref{fml:Sensitivity} to find the sensitivity.
This constitutes a lower bound on the experimentally achievable sensitivity, as we have neglected the effects of, for example, photon collection efficiency and state preparation fidelity. Because other dressed states exhibit the same overheads, however, this approximation is sufficient.

\begin{figure}[tb]
	\centering
		\includegraphics[width=\textwidth]{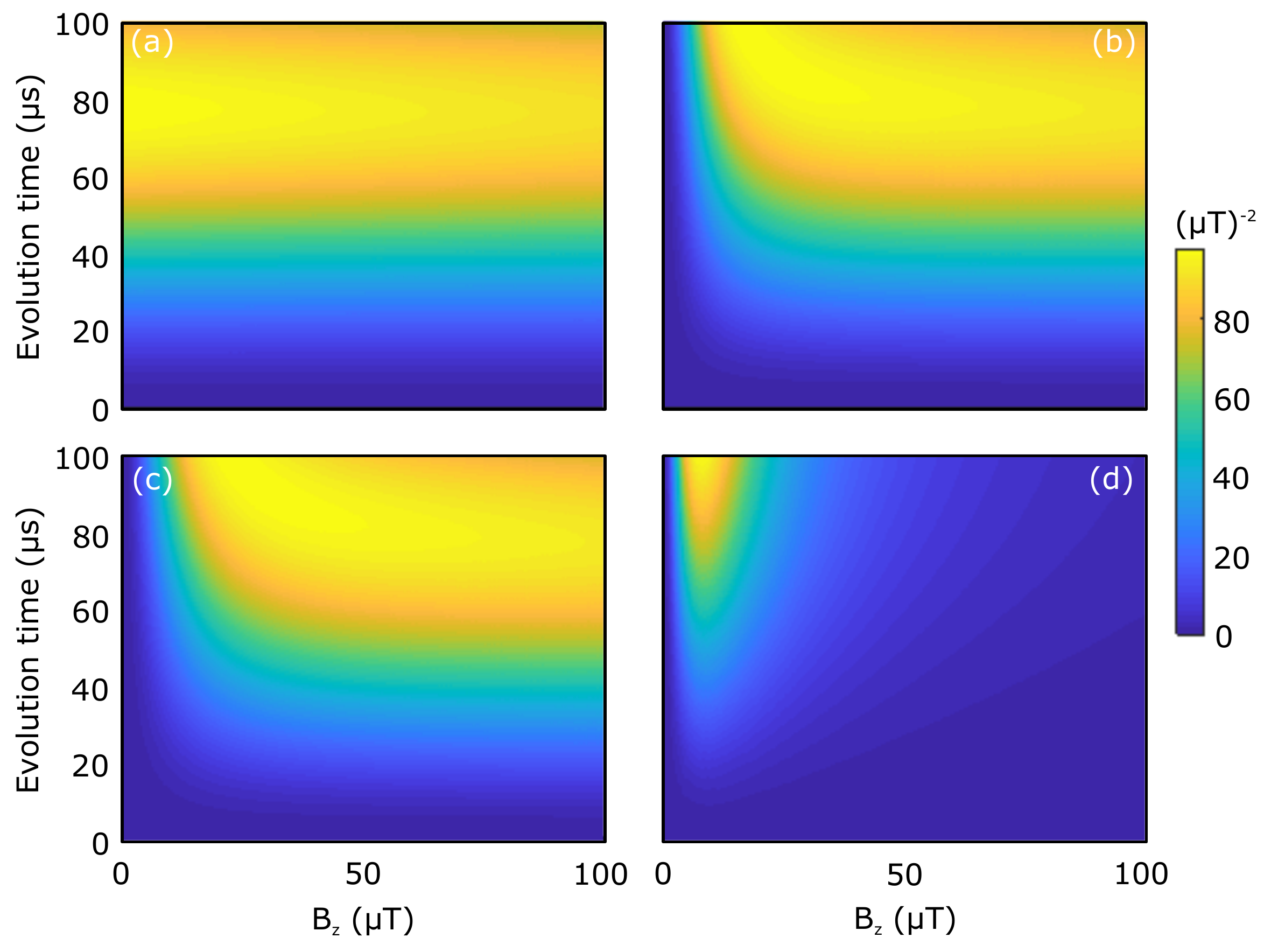}
		\caption{Quantum Fisher information neglecting decoherence. (a) Plot of $J_{\mathrm{CCI}} (B_z; t)$ at $\Phi = 0$ showing that the superposition of $\ket{ \Psi_{+1} }$ and $\ket{ \Psi_{-1} }$ is sensitive to a wide range of magnetic fields, including weak fields. (b) For $\Phi = \pi / 2$ (the value of $\Phi$ used to prepare a single dressed state in the main text), $J_\mathrm{CCI} (B_z; t)$ drops off for low magnetic fields. (c) For $\Phi = \pi / 4$ (the value of $\Phi$ at which the superposition state has the longest coherence time), $J_\mathrm{CCI} (B_z; t)$ appears very similar to the case of $\Phi = \pi / 2$. (d) Plot of $J_\mathrm{2MW}$ showing that the 2MW dressed states are sensitive to a much narrower range of magnetic field strengths.}
	\label{QFI}
\end{figure}

We find the QFI of our CCI dressed states by first numerically calculating the density matrix $\hat{\rho}_\mathrm{CCI}$.
After our initialization pulses to prepare our system, the Hamiltonian in Eq.\,(1) of the main text becomes time independent:
\begin{align}
\hat{\mathcal{H}}_\mathrm{CCI} = \frac{\hbar \Omega}{2} \left( \ket{ -1 } \bra{ 0 } + \ket{ +1 } \bra{ 0 }+ e^{i \Phi} \ket{ -1 } \bra{ +1 } + \mathrm{H.c.} \right) \ .
\end{align}
\noindent The general expression for the dressed states for any global phase $\Phi$ is given by
\begin{align}
\ket{ \Psi_k } = \frac{1}{\sqrt{3}} \left( e^{i (\Phi / 3 - 2 \pi k / 3)} \ket{ -1 } + \ket{ 0 } + e^{-i (\Phi / 3 - 2 \pi k / 3)} \ket{ +1 } \right) \ .
\end{align}
\noindent Following Degen\,\textit{et al.}\,\cite{DegenEtAl2017}, we use sensing states that are superpositions of the dressed states, such that our initial state is of the form $\ket{ \psi_{\mathrm{init}} } = \frac{1}{\sqrt{2}} \left( \ket{ \Psi_{k} } + \ket{ \Psi_{\ell} } \right)$,  $\lbrace k, \ell \rbrace \in \lbrace -1, 0, 1 \rbrace$, $k \neq \ell$ (i.\,e. we work within a two-dimensional subspace determined by the two dressed states $\ket{ \Psi_k }$ and $\ket{ \Psi_\ell }$).
We consider the sensing of a static magnetic field, given by the signal Hamiltonian $\hat{\mathcal{H}}_B / \hbar = 2 \pi \gamma_{\mathrm{NV}} B_z \hat{S}_z$, where $\hat{S}_z$ is the $S = 1$ spin matrix along the NV quantization axis, $B_z$ is the static magnetic field we would like to sense, and $\gamma_{\mathrm{NV}} = \SI{2.8}{MHz/G}$ is the gyromagnetic ratio of the NV.
The total Hamiltonian is therefore given by $\hat{\mathcal{H}} = \hat{\mathcal{H}}_\mathrm{CCI} + \hat{\mathcal{H}}_B$.
We numerically solve the equation of motion for the density matrix $\hat{\rho}_\mathrm{CCI}$
\begin{align}
\frac{\partial \hat{\rho}_\mathrm{CCI}}{\partial t} = - i \left[ \hat{\rho}_\mathrm{CCI}, \hat{\mathcal{H}} / \hbar \right]
\end{align}

\noindent to find $\hat{\rho}_\mathrm{CCI}$ as a function of evolution time $t$, magnetic field strength $B_z$, and global phase $\Phi$.
From $\hat{\rho}_\mathrm{CCI}$ we can calculate the QFI for the initial state $\ket{ \psi_\mathrm{init} } = \frac{1}{\sqrt{2}} \left( \ket{ \Psi_{-1} } + \ket{ \Psi_{+1} } \right)$. Figure\,\ref{QFI}(a)-(c) shows plots of $J_\mathrm{CCI} (B_z; t)$ at $\Phi = 0, \pi / 2, \pi/4$, respectively.
We find that our dressed states can sense a wide range of magnetic field strengths $B_z$, for many values of $\Phi$.
In particular, at $\Phi = 0$, $J_\mathrm{CCI} (B_z)$ remains large even for small magnetic fields.
This is a consequence of the fact that $\ket{ \Psi_{-1} }$ and $\ket{ \Psi_{+1} }$ are degenerate at $\Phi = 0$, and a magnetic field lifts this degeneracy. Note that we do not include decoherence in our calculation of $\hat{\rho}_\mathrm{CCI}$.

\begin{figure}[tb]
	\centering
		\includegraphics[width=0.65 \textwidth]{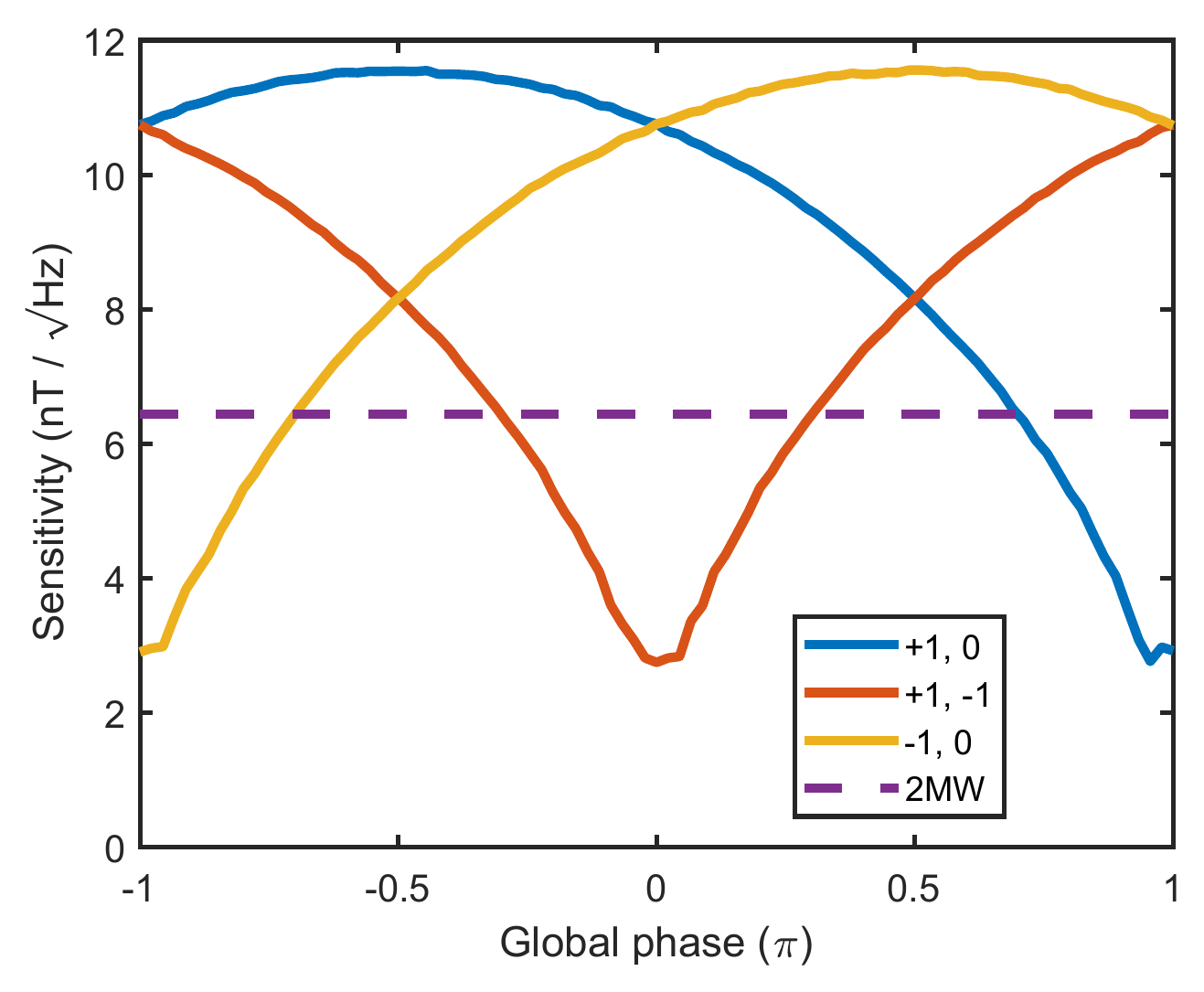}
		\caption{Best sensitivities $S_\mathrm{CCI}$ for each possible superposition of CCI dressed states. For comparison, the best sensitivity of the 2MW dressed states $S_\mathrm{2MW}$ is indicated by a dotted line.}
	\label{fig:Sensitivity}
\end{figure}

We also compare the QFI of our CCI dressed states to the 2MW dressed states, which are created using two microwave driving fields\,\cite{XuEtAl2012}, and which form the most closely comparable dressed states compared to our work. The Hamiltonian describing this system is given by
\begin{align}
\hat{\mathcal{H}}_{\mathrm{2MW}} = \frac{\hbar \Omega}{2} \left( \ket{ -1 } \bra{ 0 } + \ket{ +1 } \bra{ 0 } + \mathrm{H.c.} \right)
\end{align}
\noindent and has eigenstates and eigenenergies given by 
\begin{align}
\ket{ \varphi_\pm } &= \frac{1}{2} \left( \ket{ -1 } \pm \sqrt{2} \ket{ 0 } + \ket{ +1 } \right) \\
\ket{ \varphi_0 } &= \frac{1}{\sqrt{2}} \left( \ket{ -1 } - \ket{ +1 } \right)
\end{align}
\noindent and
\begin{align}
E_{\pm 1} = \pm \Omega / \sqrt{2} , \, E_0 = 0 \ .
\end{align}
\noindent As for the CCI dressed states, we consider a superposition of two of the 2MW dressed states: $\ket{ \psi_\mathrm{init} } = \frac{1}{\sqrt{2}} \left( \ket{ \varphi_+ } + \ket{ \varphi_0 } \right)$.
We calculate the density matrix $\hat{\rho}_\mathrm{2MW}$ in the presence of $\hat{\mathcal{H}}_\mathrm{2MW}$ and $\hat{\mathcal{H}}_B$, from which we calculate $J_\mathrm{2MW} (B_z;t)$, the QFI of the 2MW dressed states as a function of magnetic field strength $B_z$ and evolution time $t$. We find that $J_\mathrm{2MW} (B_z;t)$ is large for only a narrow range of magnetic field strengths [see Fig.\,\ref{QFI}(d)].
Defining the sensing bandwidth of the states as the FWHM of $J(B_z;T_\mathrm{coh})$, we find that our CCI dressed states can have a sensing bandwith $\sim \! 1000$ times larger than that of the 2MW dressed states for a driving strength of $\Omega / 2 \pi = \SI{500}{kHz}$, as can be anticipated from Fig.\,\ref{QFI}.
To find the sensitivity of superpositions of the CCI and 2MW dressed states, we find $J(B_\mathrm{min};T_\mathrm{coh})$ as described above.
The coherence times of the dressed states have been measured in Barfuss\,\textit{et al.} as a function of $\Phi$ \cite{BarfussEtAl2018}.
Using experimentally obtained values of $T_\mathrm{coh}^{k,\ell}$, we find the sensitivities $S_\mathrm{CCI}^{k, \ell} (\Phi)$, $\lbrace k, \ell \rbrace \in \lbrace -1, 0, +1 \rbrace$, $k \neq \ell$ and plot them in Fig.~\,\ref{fig:Sensitivity}.
For example, for the $\lbrace +1, -1 \rbrace$ superposition state, we find a sensitivity of $\approx \SI{2.75}{nT/\sqrt{Hz}}$ at $\Phi = 0$.
In each case, the superposition states are most sensitive when the two states are degenerate and have their lowest coherence times (i.\,e. at $\Phi = -\pi, 0, \pi$ for $\lbrace k, \ell \rbrace = \lbrace -1, 0 \rbrace$, $\lbrace +1, -1 \rbrace$, and $\lbrace +1, 0 \rbrace$, respectively).
We take the coherence time of the 2MW dressed states to be $\SI{5.91}{\micro s}$.
The published value is $\SI{18.9}{\micro s}$ for a drive strength of $\SI{1.6}{MHz}$; for a drive strength of $\SI{500}{kHz}$, as we use in our measurements in Barfuss\,\textit{et al.}, this corresponds to $T_\mathrm{coh} = \SI{5.91}{\micro s}$, assuming $T_\mathrm{coh}$ scales linearly with driving strength $\Omega$\,\cite{XuEtAl2012, BarfussEtAl2018, BarfussEtAl2015}.
We then find $S_\mathrm{2MW} = \SI{6.43}{nT/\sqrt{Hz}}$, which is indicated in Fig.~\,\ref{fig:Sensitivity} for comparison with the CCI dressed state sensitivities.

%============================ 
\section{Details of the CCD Scheme}
\label{sec:NVStruct}

The negatively charged NV centre, which we focus on in this work, possesses an $S=1$ electronic spin ground state from which the dressed states under study emerge. 
This spin system is composed of the eigenstates~$\ket{m_s}$ of the spin projection operator~$\hat{S}_z$ along the NV axis, with $m_s = 0,\pm 1$ being the corresponding spin quantum numbers [see Fig.\,1(a) of the main text].
In absence of symmetry breaking fields, spin-spin interactions split the degenerate $\ket{\pm 1}$ states from $\ket{0}$ by an energy~$h D_0$ with $D_0 =\SI{2.87}{GHz}$ and the Planck constant~$h$\,\cite{DohertyEtAl2013}.
Applying a static magnetic field $B_z$ along the NV axis lifts the degeneracy of $\ket{\pm 1}$ and causes a Zeeman splitting of an energy $2 h \gamma_{\rm{NV}} B_z$, with $\gamma_{\rm{NV}} = \SI{2.8}{MHz/G}$\,\cite{DohertyEtAl2013}. 
The NV spin can be readily polarised into $\ket{0}$ by optical pumping with a green laser, whereas spin-dependent fluorescence allows for optical readout of the spin state\,\cite{GruberEtAl1997}.
While the NV spin states exhibit a hyperfine splitting due to the nuclear spin of the $^{14}$N nucleus, we restrict ourselves to the nuclear spin subspace with quantum number $m_I = +1$ for experimental simplicity.

To form our CCD scheme, we apply two MW fields and one ac strain field.
While MW fields address the $\ket{0} \leftrightarrow \ket{\pm 1}$ transitions (supplied by a near-field antenna close to the sample), strain can coherently drive the nominally magnetic dipole-forbidden $\ket{-1} \leftrightarrow \ket{+1}$ transition [see Fig.\,1(a) of the main text]\,\cite{BarfussEtAl2015}. 
We realise this strain driving by placing a single NV in a mechanical resonator, which we actuate by mechanical excitation with a piezo element.
By application of an appropriate external static magnetic field, the splitting of the $\ket{\pm 1}$ states is brought into resonance with the resulting, time-varying strain field. 
With this, the strengths, relative phases and detunings of all driving fields can be individually controlled, allowing for full, coherent control of our three-level CCD\,\cite{BarfussEtAl2018}. For the ensuing state transfer protocol, we implemented arbitrary waveform control [see below] of the MW field Rabi frequencies $\Omega_{1,2}(t)$, while the Rabi frequency of the mechanical drive $\Omega$ remained constant throughout our experiments.

%============================ 
\section{Generation of arbitrary MW field pulse shapes}
\label{sec:Setup}

\begin{figure}[tb]
	\centering
		\includegraphics{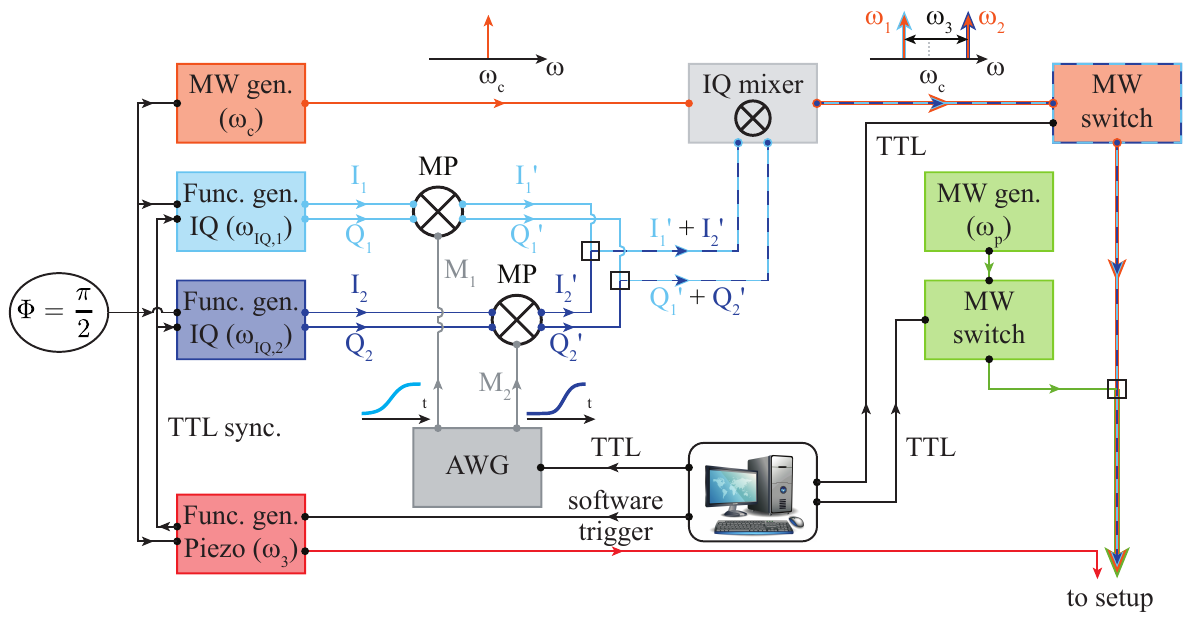}
		\caption{Setup used for arbitrary waveform control of the MW driving fields.
		The two MW driving tones with frequencies~$\omega_{1} = \omega_c - \omega_{\rm{IQ},1}$ and $\omega_{2} = \omega_c + \omega_{\rm{IQ},2}$ are created via I/Q frequency modulation of a carrier signal with frequency~$\omega_c$.
		Therefore, two pairs of ($I$,$Q$) signals with frequencies~$\omega_{\rm{IQ},1} = \SI{2.4}{MHz}$ and $\omega_{\rm{IQ},2} = \omega_3 - \omega_{\rm{IQ},1}$ are combined. Each pair exhibits an appropriate phase shift of the $I$ and $Q$ signal to suppress one modulation sideband.
		Additionally, the ($I$,$Q$) pairs' amplitudes are modified by multiplication with the signal of an arbitrary waveform generator (AWG) using a four-quadrant multiplier (MP).
		The mechanical driving field is created by function generator actuating the Piezo element near resonant with frequency~$\omega_3$.
		We establish phase-locking of the three driving fields to a global driving phase of $\Phi = \pi/2$ by pulsed output synchronisation and locking of the MW, I/Q and Piezo function generators to the same \SI{10}{MHz} reference signal.
		An additional MW probe field ($\omega_p$) is added to the MW driving fields for manipulation of the dressed spin states.}
	\label{fig1som}
\end{figure}

In our experiment, we create the pulse envelopes of the MW field amplitudes used for driving the $\ket{0} \leftrightarrow \ket{\pm 1}$ transitions through an I/Q frequency modulation technique [see Fig.\,\ref{fig1som}]. A carrier signal at frequency~$\omega_c$ is modulated with appropriate modulation signals $I$ and $Q$. Both modulation signals are composed of two individually generated pairs of ($I$,$Q$) signals at the frequencies~$\omega_{\rm{IQ},1}$ for ($I_1$, $Q_1$) and $\omega_{\rm{IQ},2}$ for ($I_2$, $Q_2$), respectively. Within each ($I$,$Q$) pair, the amplitudes are constant and equal, although the amplitudes of the pair ($I_1$, $Q_1$) differ in general from ($I_2$, $Q_2$). That is, $I_1$ and $Q_1$ differ from each other only by a phase shift. This phase shift, which is $\phi_{\rm{IQ},1} = -\pi/2$ and $\phi_{\rm{IQ},2} = \pi/2$, allows for the suppression of one modulation sideband for each ($I$,$Q$) signal pair. After combining both modulation pairs, the resulting modulated frequency spectrum contains two frequency components, namely $\omega_{1} = \omega_c - \omega_{\rm{IQ},1}$ and $\omega_{2} = \omega_c + \omega_{\rm{IQ},2}$. Modifying each ($I$,$Q$) pair's amplitude with well-defined envelope functions using an AWG and a voltage multiplier then enables us to arbitrarily shape the amplitudes of both MW driving fields. \\

In order to achieve phase-locking between the driving fields, the MW source (Standford Research Systems, SG384) and the function generators supplying the Piezo actuation and the ($I$,$Q$) signals (Keysight, 33522B) are connected to the same \SI{10}{MHz} reference signal. To set the global driving phase to $\Phi = \pi/2$, the output of the Piezo function generator is triggered via a software command. After receiving a trigger pulse, a subsequent trigger is forwarded to the ($I$,$Q$) signal generators to start their outputs. \\

The ($I$,$Q$) signals' envelopes are synthesised by an arbitrary waveform generator (Tektronix, AWG 5014C), whose signals modifies the ($I$,$Q$) pairs' amplitudes via four-quadrant multiplication (Analog Devices, AD734) before both ($I$,$Q$) pairs are combined (MiniCircuits, ZFSC-2-6+). For ($I$,$Q$) modulation we use the in-built ($I$,$Q$) modulator of our MW source. \\

An additional MW tone used for manipulation  (Rhode \& Schwarz, SMB 100A) is added to the MW driving fields via a MW combiner (MiniCircuits, ZFRSC-42-S+). All MW pulses are controlled via digital pulses from a fast pulse generator card (SpinCore, PBESR-PRO-500), which triggers the AWG and the MW switches (MiniCircuits, ZASWA-2-50DR+). \\

Creating the MW pulses in this way is limited by the vertical resolution of the four-quadrant multiplier (MP), which exhibits a noise spectral density of \SI{1}{\micro V/\sqrt{Hz}}. Hence, the estimated noise amplitude within the \SI{10}{MHz} bandwidth of the MP is \SI{3.2}{mV_{rms}}. The finite jumps at the beginning and the end of our driving field ramps are determined by the factor $\varepsilon$ defined in the main text scaled with the maximum output of the AWG, given by \SI{4.5}{V}. As additional noise is added within the ($I$,$Q$) modulation, choosing $\varepsilon = \SI{e-3}{}$ yields a discontinuity step that is comparable to the noise amplitude of our MW signals.

% FIGURE SOM

\end{document}